\documentclass[%
 aip,
rsi,%
 amsmath,amssymb,
 reprint,%
]{revtex4-1}

\usepackage{graphicx}
\usepackage{dcolumn}
\usepackage{bm}
\usepackage{pifont}
\usepackage{wasysym}
\usepackage[dvipsnames]{xcolor}
\usepackage{multirow}

\begin{document}

\preprint{AIP/123-QED}

\title[]{High Quality 3-Dimensional Aluminum Microwave Cavities}

\author{M. Kudra}
\email{kudra@chalmers.se, per.delsing@chalmers.se}

\author{J. Bizn\'{a}rov\'{a}}
\author{A. F. Roudsari} 
\affiliation{ 
Department of Microtechnology and Nanoscience, Chalmers University of Technology, Gothenburg, Sweden
}%
\author{J. J. Burnett}
\affiliation{ 
National Physical Laboratory, Hampton road, Teddington, TW11 0LW, UK
}%
\author{D. Niepce}
\author{S. Gasparinetti}
\affiliation{ 
Department of Microtechnology and Nanoscience, Chalmers University of Technology, Gothenburg, Sweden
}%
\author{B. Wickman}
\affiliation{Department of Physics, Chalmers University of Technology, Gothenburg, Sweden
}%
\author{P. Delsing}
\email{kudra@chalmers.se, per.delsing@chalmers.se}
\affiliation{ 
Department of Microtechnology and Nanoscience, Chalmers University of Technology, Gothenburg, Sweden
}%

\date{\today}

\begin{abstract}
We present a comprehensive study of internal quality factors in superconducting stub-geometry 3-dimensional cavities made of aluminum. We use wet etching, annealing and electrochemichal polishing to improve the as machined quality factor. We find that the dominant loss channel is split between two-level system loss and an unknown source with 60:40 proportion. A total of 17 cavities of different purity, resonance frequency and size were studied. Our treatment results in reproducible cavities, with ten of them showing internal quality factors above 80 million at a power corresponding to an average of a single photon in the cavity. The best cavity has an internal quality factor of 115 million at single photon level.
%
\end{abstract}

\maketitle
Quantum information with superconducting circuits is a 
leading platform for realizing a practical quantum computer. One promising approach is to encode the information in harmonic oscillators\cite{Burnett_aluminium_planar_res_2018,Kirchmair_2017,Reagor_10_ms,Reagor_Quantum_Memory_2016,niobium_3D_TLS}. Among different types of harmonic oscillators three-dimensional (3D) cavities have long lifetimes\cite{Reagor_10_ms,niobium_3D_TLS} and have been successfully integrated with qubits\cite{Reagor_Quantum_Memory_2016,100_photon_cat_2013,Ofek_2016,hu2019quantum_binomial_error_correction,Microwave_node_3D_cavities}. Out of different 3D cavity geometries, stub-geometry 3D cavities have been demonstrated to have millisecond lifetimes\cite{Reagor_Quantum_Memory_2016} at single photon level with strong dispersive coupling to the qubit. The lifetime of the cavity is inversely proportional to the internal quality factor. Although a recipe on how to make these cavities can be found in Reagor et al.\cite{Reagor_Quantum_Memory_2016} there is no systematic study on how the different parameters and treatments influence the internal quality factor. Here, we examine how different grades of aluminum, cavity height, cavity frequency and three different treatments influence the internal quality factor of the stub-geometry cavity. We find that by etching and annealing the cavities, their internal quality factor reproducibly exceeds 80 million. 
\begin{figure}[b!]
   \includegraphics[width=0.45\textwidth]{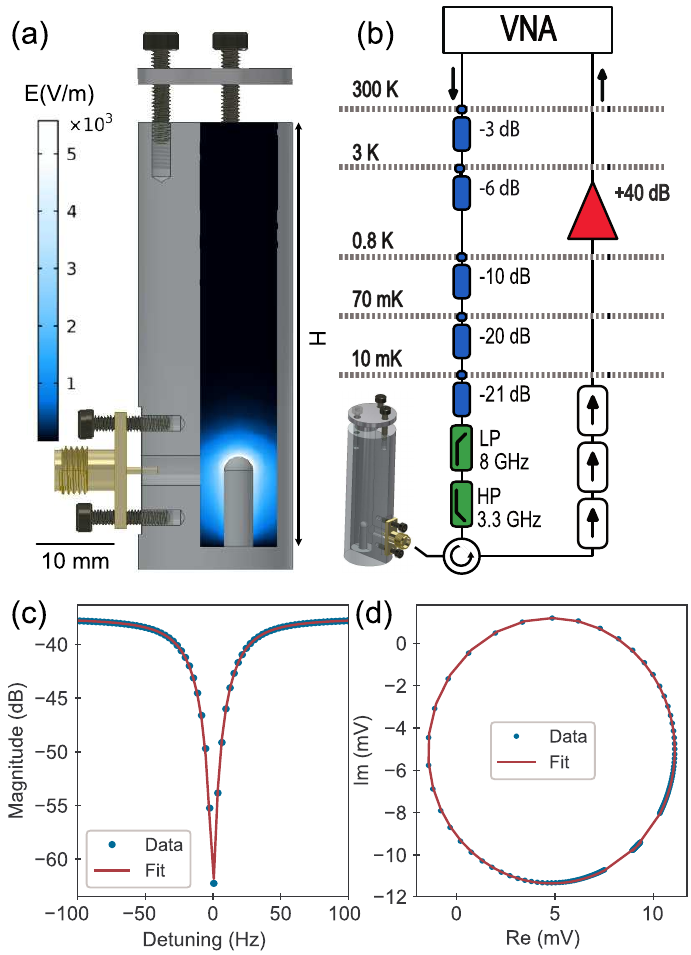}
\caption{\label{fig:setup} (a) Drawing of the stub-geometry cavity with the simulated electric field amplitude displayed in color scale. (b) Schematics of the experimental setup. Cavities are mounted at the mixing chamber of a dilution refrigerator (temperature T=10~mK) and measured in reflection with a vector network analyzer (VNA). Example of the data fitted to a circle fit\cite{Probst_2015}(c) magnitude and (d) quadratures.}
\end{figure}

A drawing of a 3D cavity and it's simulated electric field are presented in  Fig. \ref{fig:setup}(a). Compared to previous work by Reagor et al.\cite{Reagor_Quantum_Memory_2016}, we make a slight change in the geometry: we add a half sphere on top of the post, to reduce the amplitude of the electric field on the surface of the post, as shown in Fig. \ref{fig:setup}(a).

The electric and magnetic fields are concentrated around the top and the bottom of the post respectively and they decay exponentially towards the lid of the cavity. Here, we used the eigenvalue solver of COMSOL Multiphysics\textsuperscript{\textregistered} to calculate the resonance frequency, participation ratios of different loss channels\cite{Reagor_10_ms} (formulas given in the results section) and to optimize the diameter of the post with respect to the diameter of the cavity so that the participation ratio of the electric field is minimal at the surfaces. The resonance frequency of the stub geometry cavity is approximately defined by the length of the post as a quarter-wave resonator. 
 
 We study the effects of the following treatments on the internal quality factor. The first treatment is etching twice in aluminum etchant (Transene aluminum etch A) for 2~hours at 50~$^{\circ}$C to remove approximately 100~$\mu$m of aluminum\cite{Reagor_Quantum_Memory_2016}. The cavities were placed in the acid with the opening facing up to prevent bubbles that form on the surface of the aluminum getting trapped. After removing cavities from the acid bath, we rinse them in water and further clean them in acetone and then isopropyl alcohol (IPA). The second treatment is annealing\cite{3Dprinted_3Dcavity_2016} which is performed for three hours in a nitrogen atmosphere at 500~$^{\circ}$C. The warming up of the furnace is gradual and takes 1.5 hours. The cooling down to room temperature takes approximately 4-5 hours. The third treatment is electrochemical polishing in a solution of phosphoric and sulphuric acid, with ratio of 60:40. The results of a study on roughness of aluminum samples\cite{Janka_Master} inspired us to electrochemicaly polish the cavities. The cavities are connected to the positive electrode of the voltage source; while a graphite rod is placed just above the central pin of the cavity as the cathode. We perform cycles of voltage sweeps from 0-15~V at a rate of 50~mV/s for about 1 hour at 30~$^{\circ}$C. Next we increase the temperature to 60~$^{\circ}$C and we continue to sweep the voltage for another half an hour. This is followed by three cycles of sweeping the voltage at 50~mV/s until the current plateau that is characteristic for the diffusion-limited electropolishing regime\cite{landolt1987Polishing} appears in the I-V curve (3-6V depending on the cavity), whereupon the voltage is held constant at the plateau for 20 min. After electrochemical polishing we rinse the cavities in water and clean them in acetone and then IPA in the same way as after etching.

After each treatment we mounted the cavities to the mixing chamber of a dilution refrigerator inside a cryoperm shield, with no magnetic components inside the shield. We measured the cavities in reflection using cryogenic circulators and isolators Fig.~\ref{fig:setup}(b). We used the so-called circle fit\cite{Probst_2015} to simultaneously fit both quadratures in the IQ plane: 
\begin{equation}
    S_{11}=ae^{i\alpha}e^{-i2\pi f\tau}\left(\frac{2Q_l/Q_c e^{i\phi}}{1+2iQ_l(f/f_r-1)}-1\right)
\end{equation}
We fit the loaded ($Q_l$) and external ($Q_c$) quality factors as well as the resonance frequency $f_r$. We then extract internal quality factor ($1/Q_i=1/Q_l-1/Q_c$). We also fit the measurement setup parameters $a$, $\alpha$ and $\tau$, where $a$ is the background offset accounting for net attenuation of the signal sent from the vector network analyzer (Fig. \ref{fig:setup}(b)), $\alpha$ is a global phase offset and $\tau$ is electrical delay in our lines. An example of the fit is given in Fig. \ref{fig:setup}(c) and (d).
\begin{table}
\caption{\label{tab:table1}Measured single photon internal quality factors of 17 aluminum cavities, either after etching and annealing (cavities $C_{1}-C_{14}$ and $C_{16}$) or after etching (cavities $C_{15}$ and $C_{17}$). }
\begin{ruledtabular}
\begin{tabular}{ccccccc}
\multirow{2}{*}{Symbol}&\multirow{2}{*}{Cavity}& \multirow{2}{*}{Material}&  Height& $f_r$& $Q_i$& $\tau_{int}$\\
& & & (mm)&(GHz)& ($10^6$)& (ms)\\
\hline
$\CIRCLE$&$C_1$ & 5N & 35 & 7.431 & 83 & 1.79\\
\includegraphics{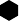}&$C_2$ & 5N & 40 & 7.417 & 66 & 1.43\\
\ding{54}&$C_3$ & 5N & 45 & 7.425 & 81 & 1.75\\
$\blacktriangledown$&$C_4$ & 5N & 50 & 7.417 & 93 & 2.02\\
$\blacktriangle$&$C_5$ & 5N & 50 & 7.427 & 79 & 1.71\\
$\blacktriangleleft$&$C_6$ & 5N & 50 & 7.428 & 91 & 1.96\\
$\blacktriangleright$&$C_7$ & 5N & 50 & 7.427 & 82 & 1.77\\
\ding{110}&$C_8$ & 5N & 50 & 6.476 & 86 & 2.11\\
\includegraphics{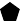}&$C_9$ & 5N & 50 & 5.478 & 94 & 2.74\\
\ding{117}&$C_{10}$ & 5N & 50 & 4.501 & 115 & 4.09\\
\ding{58}&$C_{11}$ & 4N & 35 & 5.932 & 30 & 0.81\\  
$\blacklozenge$&$C_{12}$ & 4N & 50 & 7.437 & 101 & 2.16\\
\includegraphics{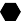}&$C_{13}$ & 4N & 50 & 5.928 & 75 & 2.03\\
\ding{72}&$C_{14}$ & 4N & 50 & 5.923 & 84 & 2.27\\ 
-&$C_{15}$ & 6081 & 35 & 5.976 & 5 & 0.13\\
-&$C_{16}$ & 6081 & 50 & 5.929 & 12 & 0.34\\
-&$C_{17}$ & 6081 & 50 & 5.939 & 7 & 0.21\\
\end{tabular}
\end{ruledtabular}
\end{table}

\begin{figure}
   \includegraphics[width=0.49\textwidth]{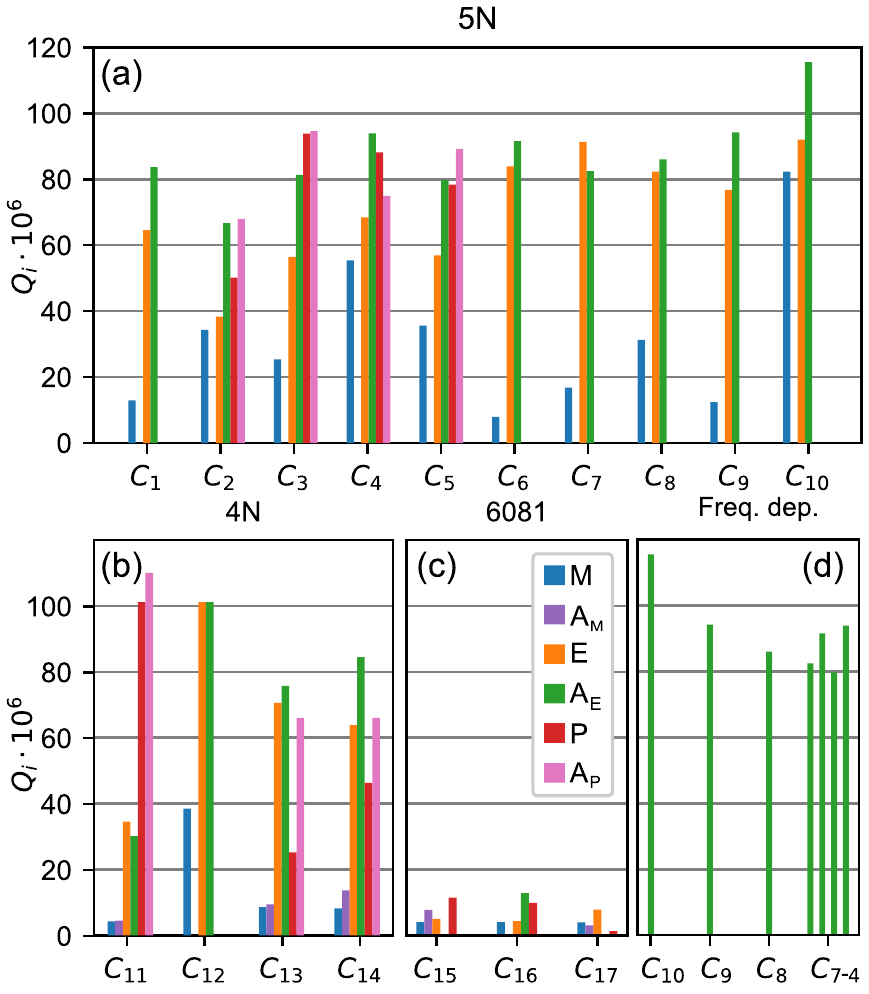}
\caption{\label{fig:bar} Single photon internal quality factor $Q_i$ of the cavities after each treatment, for cavities made out of (a) 5N, (b) 4N and (c) 6081 aluminum. M-after machining, $\mathrm{A}_\mathrm{M}$-after annealing (applied to cavities $C_{11},C_{13-15}$ and $C_{17}$), E-after etching, $\mathrm{A}_\mathrm{E}$-after etching and annealing, P-after etching, annealing and electrochemical polishing and $\mathrm{A}_\mathrm{P}$-after etching, annealing, electrochemical polishing and a second annealing step. (d) Dependence of single photon internal quality factor on resonance frequency after etching and annealing. Cavities $C_{10-4}$ are made from 5N aluminum with resonance frequencies ranging from 4.5~GHz ($C_{10}$) to 7.5~GHz ($C_{7-4}$). See Table \ref{tab:table1} for exact frequencies.}
\end{figure}

\begin{figure}
   \includegraphics[width=0.47\textwidth]{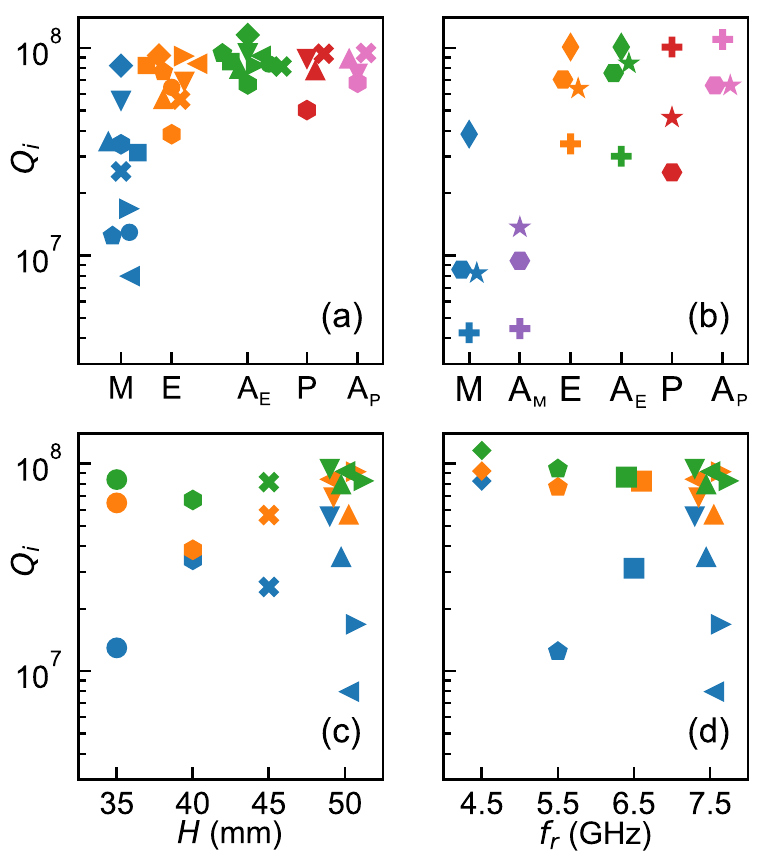}
\caption{\label{fig:breakout} (a,b) Single photon internal quality factor as a function of the treatment applied for the cavities made from (a) 5N (99.999\%) and (b) 4N (99.99\%) aluminum. Dependence of internal quality factor on the (c) height of the cavity (cavities $C_{1-7}$) and (d) resonance frequency (cavities $C_{4-10}$). Individual cavities are coded with symbols listed in Table \ref{tab:table1} and treatments are coded with same colors as in Fig. \ref{fig:bar}. The treatments were applied sequentially in order presented in (a) for 5N cavities and (b) for 4N cavities.}
\end{figure}

The results are summarized in Table~\ref{tab:table1} and Fig.~\ref{fig:bar}. Each cavity is represented by a symbol and given a name ($C_{1-17}$). The material, height and resonance frequency of each cavity are listed in the table. The internal quality factor of the cavities at single photon level is shown in Figs. \ref{fig:bar} and \ref{fig:breakout}, where each treatment is represented by a color. For the results presented in Fig. \ref{fig:bar}(a), (b) and (c), the order of the colored bars from left to right is based on the treatment sequence that the as-machined cavity has received. The cavities may have undergone some, or all of the treatments. Cavities $C_{15-17}$ (Fig. \ref{fig:bar}(c)) were machined out of an aluminum alloy 6081 which contains between 96 to 98\% aluminum and the rest is mostly magnesium, silicon and manganese. Regardless of the treatment, the internal quality factor of these cavities did not improve significantly, and the best quality factor was about 12 million. We thus conclude that quality factors of cavities $C_{15-17}$ is limited by the impurities in the material.

Cavities made from 5N (99.999\%) aluminum ($C_{1-10}$, Figs. \ref{fig:bar}, \ref{fig:breakout} and Table \ref{tab:table1}) are measured as-machined, after etching and then the etched cavities were annealed (Fig. \ref{fig:breakout}(a)). After machining, there is a wide spread of quality factors from 8 all the way to 82 million, with an average quality factor of 31 million. We attribute this to defects caused by machining and possible impurities that can be introduced. 

After etching, the quality factor increased for all of the cavities, leading to an average quality factor of 71 million with a spread of 54 million between the maximum and minimum quality factors. Etching around 100 $\mu$m of aluminum of the cavity surface seems to remove most of the machining defects. However, after etching, aluminum oxide forms on the surface in clean-room ambient atmosphere in an uncontrolled manner, therefore it can cause some spread in the results. During the annealing process, the increased mobility of the atoms allows for restoring the defects in the aluminum lattice and the oxide and the interface of the two. The average quality factor after annealing is 88 million, and the spread in quality factor is reduced to 49 million. Cavities $C_{4-7}$ are nominally identical. The internal quality factors of these four cavities are within 8\% of their average value (Table \ref{tab:table1}).

In Fig. \ref{fig:breakout}(a) and (b) we compare the performance of the cavities made from 5N (99.999\%) aluminum ($C_{1-10}$) and 4N (99.99\%) aluminum ($C_{11-14}$). While it is not possible to predict which as-machined cavity will have a better quality factor after the treatments, both types show an average quality factor of above 80 million after etching and annealing (green). Using the higher purity (5N) aluminum would not give us any leverage unless the more dominant sources of loss (discussed later), are eliminated.

A total of seven cavities made from 5N ($C_{2-5}$) and 4N ($C_{11}$ and $C_{13-14}$) aluminum were electrochemically polished (see Figs. \ref{fig:bar} and \ref{fig:breakout} (a) and (b) in red).  Although the surface of all the cavities got a mirror like finish after the polishing step, the improvement of the internal quality factor was not conclusive. For example, cavity $C_{11}$ had a $Q_i$ of 30 million after etching and annealing and the quality factor improved to 100 million after adding the polishing step. However, compared to the etched and annealed $Q_i$, polishing the other cavities either deteriorated the $Q_i$ or just slightly improved it (see Figs. \ref{fig:bar} and \ref{fig:breakout} (a) and (b) in red). Adding an annealing step after the polishing step (see Figs. \ref{fig:bar} and \ref{fig:breakout} (a) and (b) in pink) improved the polished $Q_i$ slightly. More investigations are needed to make electrochemical polishing a more reliable procedure for improving of the internal quality factor.

With the above observations, we set out to determine the dominant loss mechanism. To explore the influence of seam loss, we made cavities $C_{1-7}$ of varying heights (35 - 50~mm, see Fig. \ref{fig:setup}(a)). The participation of the seam loss exponentially decays with the height of the cylindrical waveguide section\cite{Reagor_Quantum_Memory_2016} so we would expect to see this trend if seam loss was the limiting loss factor. No such trend is visible (Fig. \ref{fig:breakout}(c)). 
After etching and annealing, regardless of the height, all of the cavities show similar $Q_i$. Therefore, we conclude that seam loss is not the limiting factor for cavities longer than 35~mm at this quality factor level. 

Next, we investigate the influence of resonance frequency on the internal quality factor. The resonance frequency of the cavities $C_{4-10}$ ranges from 4.5~-~7.5~GHz. Here the length of the center pin was changed to get the desired resonance frequency. The length of the waveguide was simultaneously adjusted so that the seam loss has the same participation ratio for all of them. For the cavities to meet these two criteria the total height of the cavity from the bottom of the pin to the lid should be kept fixed (H in Fig. \ref{fig:setup}(a)). There is a slight trend of lower frequencies having higher internal quality factor (Figs. \ref{fig:breakout} and \ref{fig:bar}(d) and Table \ref{tab:table1}). This could be attributed to the lower density of states of the TLSs at lower frequencies\cite{skacel2015probingDOSofTLS}.
\begin{figure*}
\includegraphics[width=0.95\textwidth]{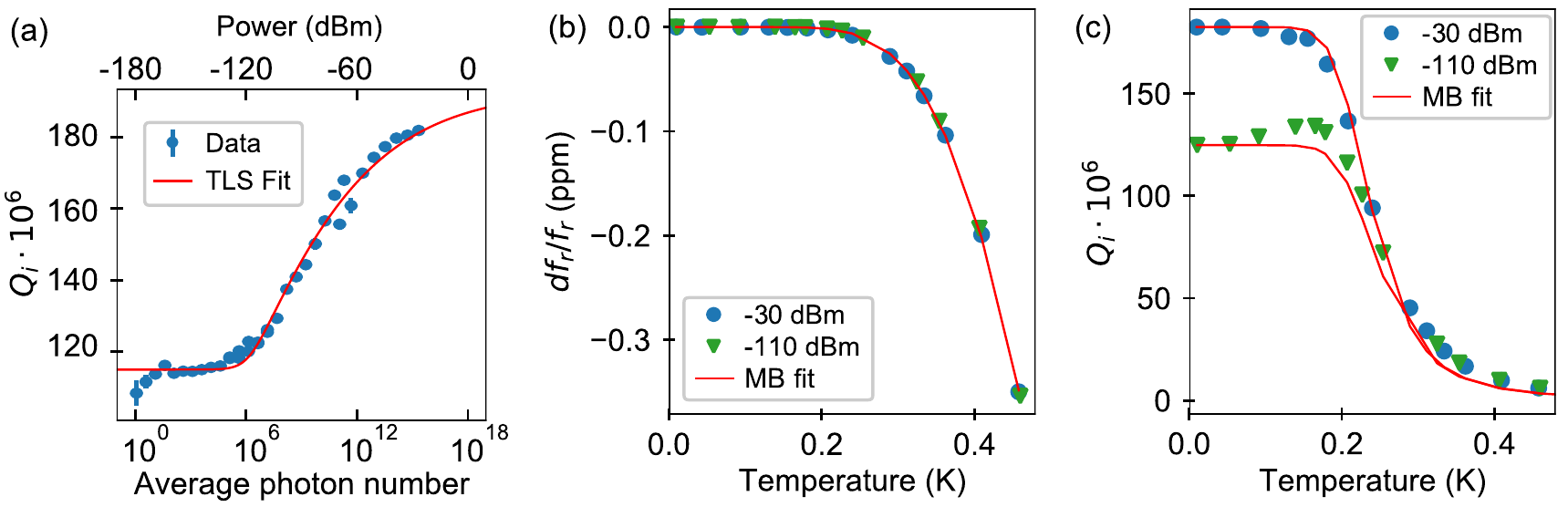}
\caption{\label{fig:TLS_and_BCS}(a) Internal quality factor as a function of average number of photons in the $C_{10}$ cavity (after etching and annealing) fitted to the TLS model (Eq.~\ref{eq_TLS}). The top axis shows the estimated power sent to the cavity. (b,c) The temperature dependence of (b) frequency and (c) internal quality factor for two different input powers (symbols) and the Mattes-Bardeen fit\cite{Gao_2008} (solid lines). Frequency and internal quality factor are fitted simultaneously using two fitting parameters, critical temperature $T_c$ and kinetic inductance ratio $\alpha$.}
\end{figure*}

The dependence of $Q_i$ on power and temperature for cavity $C_{10}$ after etching and annealing are presented in Fig. \ref{fig:TLS_and_BCS}. We fitted the internal quality factor ($Q_i$) as a function of average number of photons to a standard TLS model\cite{TLS_fluctuations_Burnett,TLS_fluctuations_Klimov,TLS_fluctuations_Schlor,TLS_Martinis} (Fig. \ref{fig:TLS_and_BCS}(a)):
\begin{equation}
    1/Q_i=F\delta_{TLS}\tanh\left(\frac{hf_r}{2kT}\right)(1+n/n_c)^{-\beta}+\frac{1}{Q_{res}}
    \label{eq_TLS}
\end{equation}
The average number of photons $n$ is estimated using the formula $n=4Q_l^2P/(Q_c\hbar(2\pi f_r)^2)$, where the power sent to the cavity $P$ is estimated from the value at the source and the known line attenuation. The first parameter we extract from the fit is the product of participation ratio of the electric field ($F$) and the two level system loss tangent ($\delta_{TLS}$). We simulated the participation ratio of the electric field $F=t_{ox}\iint{|E|^2dS}/\epsilon_r\iiint{|E|^2dV}=3.6\cdot10^{-7}$ assuming the thickness of the oxide layer $t_{ox}=$~5~nm and the relative permittivity of aluminum oxide  $\epsilon_r=$~10. This yields the loss tangent of the surface oxide of $\delta_{TLS}=9.8\cdot10^{-3}$. The second fit parameter is the critical photon number $n_c=1.3\cdot10^{6}$. It is the number of photons that produce an electric field strength that can saturate a TLS. The small participation ratio that we simulated is confirmed in the high critical photon number. The critical photon number is close to unity in planar and microstrip technology\cite{Burnett_aluminium_planar_res_2018,Kirchmair_2017}. The third fitting parameter is the quality factor of the residual loss mechanism, $Q_{res}=193\cdot10^{6}$. Finally $\beta=0.11$ is a fitting parameter usually found to be around 0.2 in both planar\cite{macha2010TLSlosses,burnett2017TLSlow} and 3D resonators\cite{niobium_3D_TLS}. Given the  difference between the high power and the low power internal quality factors, we calculate that the TLSs contribute to 63\% of the total loss in the resonator at low power. We are not able to present data for higher photon numbers since the resonator becomes non-linear; the resonance frequency shifts to lower frequencies and the line shape is no longer reliably fitted by the circle fit routine. The other cavities show similar dependence of internal quality factor to power.  

By sweeping the temperature, we can probe the sensitivity of the cavity to thermally excited quasiparticles. In Fig.~\ref{fig:TLS_and_BCS}(b) and (c) we show a Mattis-Bardeen fit\cite{Gao_2008} of the frequency and the internal quality factor as a function of temperature. The fit is in the clean limit for a bulk superconductor\cite{Gao_2008} and it has two fitting parameters. The first one is the bulk aluminum critical temperature $T_c$=1.18~K, which is quite close to the literature value\cite{popel1989electromagnetic}. The second one is the kinetic inductance ratio $\alpha=5.07\cdot10^{-5}$. $\alpha$ is given by $\lambda\cdot p_H$ where $p_H=\iint{|H|^2dS}/\iiint{|H|^2dV}$ is simulated to be $p_H=814.8$~m$^{-1}$. Thus we can extract the effective penetration depth $\lambda=62$~nm, which compares relatively well to the textbook values\cite{Gao_2008}. Both frequency dependence and the internal quality factor dependence were fitted simultaneously with the same critical temperature and kinetic inductance. The slight increase in internal quality factor between 10 mK and 200 mK for the low power trace (green triangles in Fig. \ref{fig:TLS_and_BCS}(c)) is due to the tanh dependence of the TLS given by Eq.~2\cite{niobium_3D_TLS}. If we assume that the $Q$-value at high power, $Q_{res}$, was caused by quasiparticles, the equivalent temperature of the quasiparticles would be 223~mK. We find it unlikely that the cavity is that hot or that the non-equilibrium density of the quasiparticles is that high, and we therefore suppose that $Q_{res}$ is due to some other unknown loss.

In conclusion, we have demonstrated that etching and annealing 3D cavities results in a reproducible recipe to make aluminum  cavities with internal quality factors at single photon level exceeding 80~million. Electrochemical polishing improved quality factors of some cavities but reduced the quality factors of others. More research is needed to make this process more reproducible. Once the total height of the cavity exceeds 35~mm, seam loss is not a limiting factor. TLS loss contributes to around 60\% of the total loss. The difference between 4N and 5N cavities is not visible at internal quality factors around 100 million.

\begin{acknowledgments}
The authors thank Mats Myremark and Lars J\"{o}nsson  for machining the cavities. This work was supported by Knut and Alice Wallenberg foundation via the Wallenberg Center for Quantum Technology (WACQT) and by the Swedish Research Council. The authors acknowledge the use of Chalmers Myfab for the treatment of the cavities. J.J.B. acknowledges financial support from the Industrial Strategy Challenge Fund Metrology Fellowship as part of the UK government's Department for Business, Energy and Industrial Strategy.
\end{acknowledgments}

\bibliography{ref}

\end{document}